# Analysis of a three phase induction motor directly from Maxwell's equations

Shayak Bhattacharjee

Department of Physics, Indian Institute of Technology, Kanpur – 208016, Uttar Pradesh, India

## Abstract

The torque developed in a three phase AC squirrel cage motor is usually expressed in terms of resistances and reactances of the stator, the rotor, and the motor as a whole. We use Maxwell's equations to find the torque in terms of geometrical parameters. This allows us to estimate the torque developed by a motor without knowing the details of its circuitry.

## I. Introduction

The analysis of the induction motor is generally the domain of electrical engineering. In this paper we present an alternative analysis based on physical considerations. Our results are in terms of geometrical parameters of the motor rather than in terms of reactances of the different components.

The standard procedure[1] for evaluating the performance of a three-phase AC squirrel cage motor is a perphase analysis of a circuit containing Thevenin equivalents of the stator, the rotor, and the load. In other words, each phase of the circuit is analysed separately and then the results for the three phases are combined. The analysis proceeds as for a transformer, with the coupling between windings taken to be a function of the rotation speed. The circuit diagram is shown in Fig. 1.

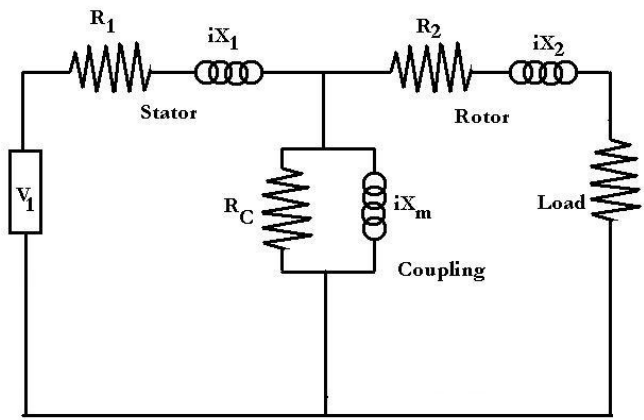

Figure 1: Circuit diagram of a three-phase squirrel cage AC motor showing the Thevenin equivalent circuit.

Here $R_1$ and $X_1$ refer to the resistance and the reactance of the stator, $R_2$ and $X_2$ to those of the rotor, and $R_{eq}$ and $X_{eq}$ are the equivalent parameters for the motor as a whole. $V_1$ is the voltage applied to the stator. From Thevenin's theorem, we have

$$R_{eq} + iX_{eq} = (R_1 + iX_1) + R_C \| iX_m, \qquad (1)$$

where $R_1 || R_2$ denotes the equivalent resistances of the parallel combination of $R_1$ and $R_2$, and

$$\tilde{V}_{eq} = V_1 \frac{R_C || iX_m}{R_1 + iX_1 + (R_C || iX_m)}. \qquad (2)$$

The torque $\Gamma$ for the equivalent Thevenin circuit is given by

$$\frac{3\frac{R_2}{s}|\tilde{V}_{eq}|^2}{[(R_{eq} + \frac{R_2}{s})^2 + (X_{eq} + X_2)^2]f_1}, \qquad (3)$$

where $f_1$ is the frequency at which the voltage is applied to the stator, $f_2$ is the angular frequency of the rotor and $s = \frac{f_2 - f_1}{f_1}$. Thus we see that the torque is expressed in terms of the resistances and reactances of the various components.

In this paper we will derive an expression for $\Gamma$ based entirely on Maxwell's equations. Our derivation requires no further knowledge of electromagnetism than can be acquired from a standard undergraduate text.[2]

## II. DERIVATION

We show the top and front views of a squirrel cage motor in Fig. 2.

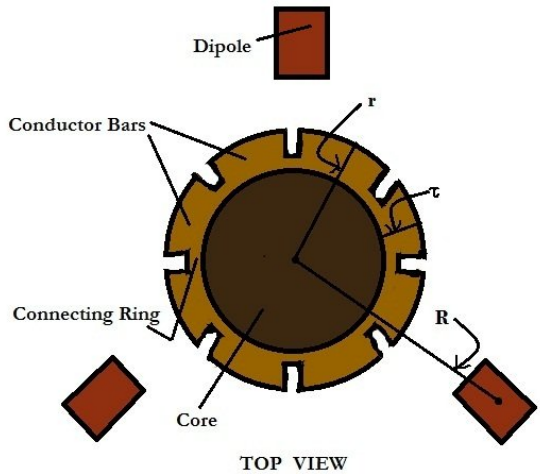

Figure 2: (a) Top view of a three-phase squirrel cage AC motor. The cage of the motor has a radius $r$, the radial extension of each conducting bar is $\tau$, and $R (>r)$ is the distance between the center of the core and the center of a dipole.

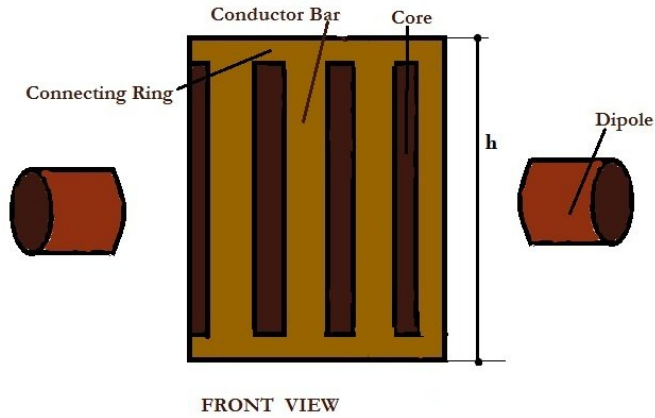

b) Front view of the motor. The height of the cage is h.

The three magnetic dipoles, each assumed to be ideal, are mounted in the horizontal plane at equal angular spacing at the radius R. The cage has radius r and height h, the conductor bars have thickness τ in the radial direction, and conductivity σ. The conductor is arranged in the form of vertical bars instead of as a continuous cylindrical shell for the purpose of constraining eddy currents in all directions other than the vertical. As we shall see, it is only the vertical eddies that contribute to the torque. The bars are placed close together so that we can treat the conductor as a continuous cylindrical shell without appreciable error. The orientations of the dipoles are shown in Fig. 3. Three phase alternating voltage at frequency Ω is now supplied to the motor, one phase to each dipole. Hence the currents through dipoles 1, 2, and 3 are $I_0 \sin \Omega t$, $I_0 \sin (\Omega t + 2\pi/3)$ and $I_0 \sin (\Omega t + 4\pi/3)$, respectively.

We now consider a point on a circle of radius r at an angle θ from a reference line, which, without loss of generality we may select as the line joining the centre of the cage to dipole 1. We show a schematic diagram in Fig. 3.

The expression for the field of a magnetic dipole is cumbersome, and therefore we use the vector potential. The vector potential of a magnetic dipole of moment m is

$$\vec{A} = \frac{\mu_0}{4\pi} \frac{\vec{m} \times \vec{r}}{|\vec{r}|^3},$$ (4)

where r is the position vector of the field point P (see Fig. 3) relative to the dipole. In our case the dipoles are oriented with their axes in the x-y plane, and A is in the z direction at all points in the x-y plane. For simplicity, we use a two-dimensional geometry.

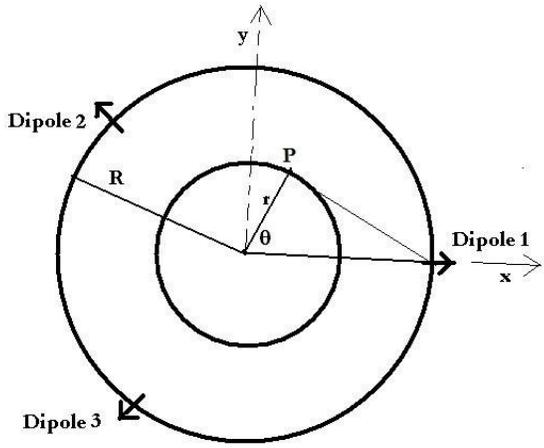

Figure 3: Schematic diagram of the AC motor. The magnetic field is calculated at a point P on the circumference of the cage.

For a dipole 1 (which has a strength M), we have
$$\vec{r} = -R\hat{i} + (r\cos\theta\hat{i} + r\sin\theta\hat{j})$$
(5a)

$$\vec{M} = M\hat{i}$$
(5b)

We define
$$\vec{A}' = \frac{\vec{A}}{\mu_0/4\pi},$$
(6)

and obtain
$$\vec{A}' = \frac{Mr\sin\theta}{(R^2 + r^2 - 2Rr\cos\theta)^{3/2}}\hat{k}.$$
(7)

where i, j and k are the unit vectors in the x, y and z directions respectively. Likewise for dipole 2,
$$\vec{A}' = \frac{-\tfrac{1}{2}Mr\sin\theta - \tfrac{\sqrt{3}}{2}Mr\cos\theta}{[R^2 + r^2 - 2Rr\cos(\theta - 2\pi/3)]^{3/2}}\hat{k},$$
(8)

and for dipole 3,
$$\vec{A}' = \frac{-\tfrac{1}{2}Mr\sin\theta + \tfrac{\sqrt{3}}{2}Mr\cos\theta}{[R^2 + r^2 - 2Rr\cos(\theta - 4\pi/3)]^{3/2}}\hat{k}.$$
(9)

We expand terms of the form $\frac{1}{(R^2 + r^2 - 2Rr\cos\alpha)^{3/2}}$ as $\frac{1}{R^3} + \frac{3r}{R^4}\cos\alpha + ....$ substitute the result into Eqs. (7)–(9) and combine them to obtain

$$|\vec{A}'| = \frac{1}{R^3}\begin{pmatrix} M_0 r\sin\theta - \frac{1}{2}M_1 r\sin\theta - \frac{\sqrt{3}}{2}M_1 r\cos\theta \\ -\frac{1}{2}M_2 r\sin\theta + \frac{\sqrt{3}}{2}M_2 r\cos\theta \end{pmatrix}$$

$$+\frac{3r}{R^4}(M_0 r\sin\theta\cos\theta + \ldots) \quad , \tag{10}$$

where $M_0$, $M_1$, and $M_2$ are the strengths of dipoles 1, 2, and 3 respectively.

Equation (10) has the form

$$|\vec{A}'| = \frac{1}{R^2}\left[\left(\frac{r}{R}\right)\begin{pmatrix} M_0 \sin\theta - \frac{1}{2}M_1 \sin\theta - \frac{\sqrt{3}}{2}M_1 \cos\theta \\ -\frac{1}{2}M_2 \sin\theta + \frac{\sqrt{3}}{2}M_2 \cos\theta \end{pmatrix} + \left(\frac{r}{R}\right)^2 (3M_0 \sin\theta\cos\theta + \ldots)\right] \tag{11}$$

which is a power series in r/R. We assume $r \ll R$ and retain only the first order terms.
We use the fact that the dipole moment is proportional to the current flowing through it, and let M be the peak strength of each dipole in the motor,

$$M_0 = M\sin\Omega t, \tag{12a}$$
$$M_1 = M\sin(\Omega t + 2\pi/3), \tag{12b}$$
$$M_2 = M\sin(\Omega t + 4\pi/3). \tag{12c}$$

We substitute Eq. (12) into the first term of Eq. (11) and obtain

$$\vec{A}' = \left\{\frac{rM}{R^3}\begin{bmatrix} \sin\Omega t \sin\theta \\ -\frac{1}{2}\left(-\frac{1}{2}\sin\Omega t + \frac{\sqrt{3}}{2}\cos\Omega t\right)\left(\sin\theta + \sqrt{3}\cos\theta\right) \\ -\frac{1}{2}\left(-\frac{1}{2}\sin\Omega t - \frac{\sqrt{3}}{2}\cos\Omega t\right)\left(\sin\theta - \sqrt{3}\cos\theta\right) \end{bmatrix}\right\}\hat{k} \quad, \tag{13}$$

which can be simplified to

$$\vec{A} = -\frac{3\mu_0}{8\pi}\frac{rM}{R^3}\cos(\theta + \Omega t)\hat{k} \quad. \tag{14}$$

By using $\vec{B} = \text{curl } \vec{A}$ and changing to cylindrical coordinates ρ, θ, z we obtain

$$\vec{B} = \frac{3\mu_0}{8\pi}\frac{M}{R^3}\sin(\theta + \Omega t)\hat{\rho}$$
$$+\frac{3\mu_0}{8\pi}\frac{M}{R^3}\cos(\theta + \Omega t)\hat{\theta} \quad. \tag{15}$$

Thus B is rotating with angular velocity $\Omega$. Because the induced current is in the z direction (due to the conducting bars), the $\theta$ component of B, when crossed with the current, creates a radial force which produces zero torque. Hence in the subsequent analysis, we will consider the $\rho$ component alone of the magnetic field.

Let $\omega$ be the rotation rate of the cage and define

$$B_0 = \frac{3\mu_0}{8\pi}\frac{M}{R^3}. \tag{16}$$

Qualitatively the rotating B drags the cage along with it so that $\omega$ is clockwise. We now quantify this drag using techniques similar to those found in Refs. 3–6. Consider a time $t_0$ and a point at the angle $\theta$. We use a frame moving at speed $\Omega r$ in the $-\theta$ direction. In this frame $B = B_0 \sin(\theta+\Omega t_0)$ and is constant in time. The cage moves with speed $(\Omega - \omega)r$ in the $\theta$ direction. If we transfer to the frame of the cage, we obtain the induced electric field

$$\vec{E} = \vec{v} \times \vec{B} = -B_0(\Omega-\omega)r\sin(\theta+\Omega t_0)\hat{z}. \tag{17}$$

We use $J = \sigma E$, and note that this constitutive relation is valid for small enough E which holds only for low values of $\Omega-\omega$ from Eq. (17). We also note that J can be assumed constant through an element of cross sectional area $\tau r d\theta$ ($\tau$ is the conductor thickness and $\tau << r$). We obtain the infinitesimal current

$$dI = \sigma[-B_0(\Omega-\omega)r\sin(\theta+\Omega t_0)]\tau r d\theta, \tag{18}$$

and the infinitesimal force is

$$d\vec{F} = \sigma h[-B_0^2(\Omega-\omega)r\sin^2(\theta+\Omega t_0)]\tau r d\theta \hat{\theta}, \tag{19}$$

from which the infinitesimal torque can be expressed as

$$d\vec{\Gamma} = -\sigma h r^3 \tau B_0^2(\Omega-\omega)\sin^2(\theta+\Omega t_0)d\theta \hat{z}. \tag{20}$$

Because B is rotating, the profiles of the field at two different times are identical except for a shift by an angle which becomes irrelevant when we integrate over the whole cage. Hence the torque does not vary with time so long as $\Omega$ and $\omega$ are constant, and we may evaluate it by setting $t_0 = 0$ in Eq. (20) and integrating over the whole range of $\theta$, $0 \leq \theta \leq 2\pi$. Hence

$$\vec{\Gamma} = -\pi\sigma h r^3 \tau B_0^2(\Omega-\omega)\hat{z}. \tag{21}$$

As expected, the torque is in the direction of rotation of the cage. Its magnitude is given by

$$\Gamma = \frac{9\mu_0^2 M^2 r^3 h \sigma \tau}{64\pi R^6}(\Omega-\omega). \tag{22}$$

If the cage has an iron core, then $\mu_0$ in Eq. (22) is replaced by $\mu_{eff}$ whose value is between $\mu_0$ and $\mu_{core}$. The entire $\mu_{core}$ will not act because the dipoles are mounted in air.

Thus the torque is proportional to the difference between the rotational frequency and the excitation frequency. Note that for low frequencies, the Thevenin circuit answer in Eq. (3) also yields a torque proportional to $\Omega-\omega$. If we use the condition $s << 1$ in Eq. (3) (where the slip factor $s = 1 - \omega/\Omega$), we obtain a simplified expression for the torque,

$$\Gamma = \frac{3s|V_{eq}|^2}{f_1 R_2}, \tag{23}$$

which shows that the torque depends linearly on the frequency difference, which is called the slip frequency.

The torque as a function of the slip has been experimentally measured and also numerically calculated[7] (see Fig. 4). As expected, the torque increases linearly in the low slip region.

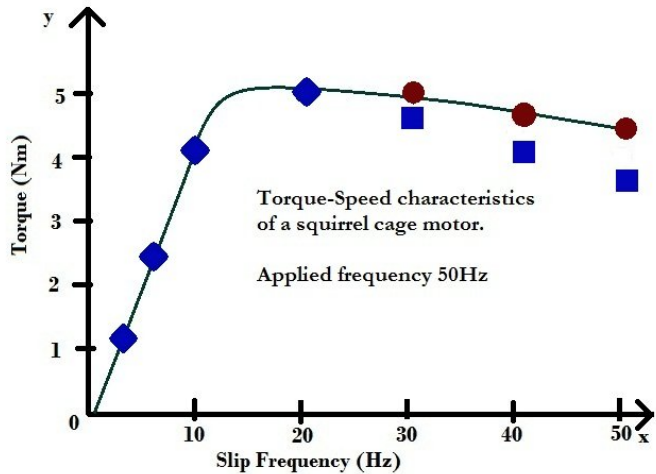

Figure 4: The torque as a function of the slip frequency. The data are taken from Ref. 7. The circles and squares correspond to experimental data and numerical calculations respectively. Points where the two have coincided exactly have been shown as diamonds.

The saturation at a slip frequency around 10 Hz is a consequence of the breakdown of linear response. As $E$ increases, the first correction to $J$ appears in the form
$J = \sigma E - \sigma'|E|^2 E$. For practical applications, the motor generally operates in the linear regime.

### III. Application

We apply Eq. (22) to a motor of type 6FXA7059 manufactured by Crompton Greaves[9] and obtain an estimate for the torque from the data and the picture in Fig. 5.

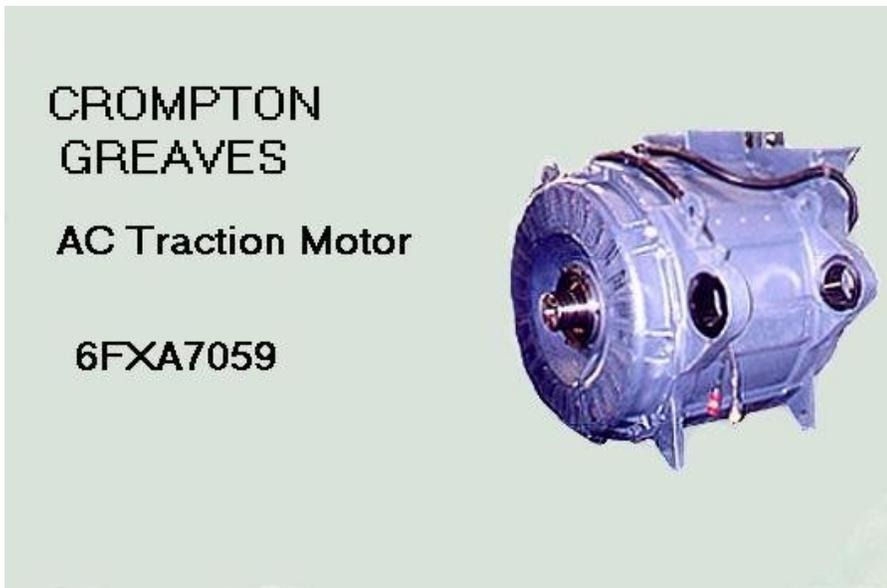

Figure 5: Crompton Greaves traction motor of type 6FXA7059.

There are (continuous ratings) six dipoles, maximum voltage is 2180 V, maximum current is 370 A; speed is 1583 rpm, and mass is 2050 kg. From Ref. 8 we see that the continuous rated torque is 6930 Nm, and that the torque increases to 10000 Nm when the current is increased to 450 A, which are the maximum ratings.

We note from Fig. 5 that $h/R = 2$ and $R/r = 3$. The current value indicates a copper wire of gauge zero zero. For equal weights of the stator and rotor there is approximately 240 m of wire in each dipole. If we assume a density of 6000 kg m$^{-3}$ for the laminated iron cage, we obtain $r = 0.16$ m, $R = 0.5$ m, and $h = 1$ m. Another assumption is that the dipoles are squares of side 20 cm so that the six dipoles cover a third of the circumference of the cage. Each dipole has about 360 turns of wire, and the dipolar strength is about 1450 A m$^2$. Taking $\mu_{eff}$ to be 100 and considering 4% slip operation at 25 Hz, the torque is about 8000 Nm. Thus we see this derivation has produced a good estimate for the torque, starting from information which would not have yielded an answer from the Thevenin formulation.

We conclude by noting an important feature of our result. The dipole moment M in Eq. (5) is proportional to the input current, and hence the torque as given by Eq. (22) depends quadratically on the input current. This result is not contained in the Thevenin circuit analysis. We have looked for data with which we could test this result. In Ref. 8 we find the data sets $I = 370$ A, $\Gamma = 6930$ Nm, and $I = 450$ A, $\Gamma = 10000$ Nm. We see that I changes by a factor of 1.22, and the torque changes by a factor of 1.44 which is consistent with Eq. (22).

## ACKNOWLEDGEMENT
I am grateful to KVPY, Government of India, for a fellowship.